# Signatures of Dirac fermion-mediated magnetic order

Paolo Sessi [1,§], Felix Reis [1], Thomas Bathon [1], Konstantin A. Kokh [2,3], Oleg. E.Tereshchenko [3,4], and Matthias Bode [1,5]

[1] Physikalisches Institut, Experimentelle Physik II, Universität Würzburg, Am Hubland, D-97074 Würzburg, Germany
[2] V.S. Sobolev Institute of Geology and Mineralogy, Siberian Branch, Russian Academy of Sciences, 630090 Novosibirsk, Russia
[3] Novosibirsk State University, 630090 Novosibirsk, Russia
[4] A.V. Rzanov Institute of Semiconductor Physics, Siberian Branch, Russian Academy of Sciences, 630090 Novosibirsk, Russia
[5] Wilhelm Conrad Röntgen-Center for Complex Material Systems (RCCM), Universität Würzburg, Am Hubland, D-97074 Würzburg, Germany

§ To whom correspondence should be addressed.
Email: paolo.sessi@physik.uni-wuerzburg.de

**The spin-momentum locking of topological states offers an ideal platform to explore novel magneto-electric effects. These intimately depend on the ability to manipulate the spin texture in a controlled way. Here, we combine scanning tunneling microscopy with single atoms deposition to map the evolution of topological states under the influence of different magnetic perturbations. We obtain signatures of Dirac fermion-mediated magnetic order for extremely dilute adatoms concentrations. This striking observation is found to critically depend on the single adatoms magnetic anisotropy and the position of the Fermi level. Our findings open new perspectives in spin engineering topological states at the atomic scale and pave the way to explore novel spin-related topological phenomena with promising potential for applications.**

**Introduction**

Manipulating unconventional spin textures represents a frontier towards quantum spintronics. Prominent systems showing novel properties associated with the spin degree of freedom are magnetic skyrmions[1] and the recently discovered topological insulators (TIs)[2,3]. TIs, in particular, offer an ideal platform to explore several exotic phenomena which, in addition of being relevant to both condensed matter and high energy physics[4-6], are expected to have direct applications in devices with unique functionalities[7-9]. The promise of this new class of materials is related to their unconventional electronic properties. Contrary to other "trivial" two-dimensional systems hosting massless Dirac fermions like graphene[10], topological states are robust against small perturbations and cannot be destroyed by the presence of defects and adsorbates as long as time-reversal symmetry is preserved[11,12]. Due to strong spin-orbit coupling, the spin is perpendicularly locked to the momentum, resulting in a chiral spin texture[13] which forbids backscattering[14,15]. This spin-momentum locking gives rise to electric charging of the magnetic texture offering new opportunities to explore topological magneto-electric effects impossible with other semiconductor materials. The fundamental requirement for their experimental realization is intimately related to the capability of manipulating the spin texture in a controlled way, as it can be achieved through the interaction of topological states with magnetic moments.

Although numerous studies aimed to shed light on the role played by magnetic perturbations, contradictory results have been obtained and a clear picture is still missing[16–25]. Results published so far seem to suggest that surface- and bulk-doped samples behave quite differently. Namely, an opening of the Dirac gap with the onset of magnetic order was exclusively reported for bulk-doped samples[16,18] while the Dirac cone remained intact for most studies employing surface doping[22–25]. This is in contradiction to theory which predicts that, due to the absence of scattering into the bulk, surface-confined Dirac fermions can mediate magnetic interaction over distances of several nanometers via the Ruderman-Kittel-Kasuya-Yosida (RKKY) interaction[26,27]. This means that topological states are expected to significantly alter their spin texture and eventually establish magnetic order for very dilute magnetic adatom concentrations. If experimentally verified, it would allow for atomic level control of the spin texture of topological states.

To date, the experimental study of the interaction of surface magnetic moments with topological states has predominantly been performed by angle-resolved photoemission spectroscopy (ARPES)[21-23] and x-ray magnetic circular dichroism (XMCD)[24-25]. Although providing invaluable contributions towards the understanding of the interaction of magnetic moments with topological states, these spatially averaging techniques do not give access to local phenomena. The broadening observed in the ARPES data, for example, may be indicative of an inhomogeneous surface magnetization. Indeed—even if long-range magnetic order is absent—short length- and temporal-scale spin-ordered states might still exist but cancel each other and may therefore remain undetected by spatially averaging techniques.

Here, we combine low-temperature scanning tunneling microscopy with single adatoms deposition technique to directly map the evolution of the electronic properties of topological states under the influence of different magnetic perturbations. We reveal signatures of surface magnetic order as soon as the magnetic moments couple ferromagnetically. By using different magnetic elements and coverages, we find that this striking observation crucially depends on two parameters: single adatoms magnetic anisotropy direction and energy level alignment.

**Results**

**Scattering scenario on topological insulators**

To obtain detailed information on the role played by magnetic perturbations, we employed energy-resolved Fourier-transformed (FT) quasi-particle interference (QPI) mapping. This technique makes use of the standing wave pattern generated by the coherent scattering of electronic states at surface defects, such as vacancies, adsorbates, impurities, or step edges. First employed to detect coherent scattering of conventional surface states on noble metals[28], it is a powerful technique to investigate the scattering properties of surfaces. It has been successfully applied to study high-temperature superconductors[29] as well as heavy-fermion systems[30,31]. Application of QPI to surfaces with non-degenerate spin bands has been pioneered on the Bi(110) surface[32] and has recently been extended to study spin-dependent scattering processes in Rashba-systems[33] and TIs[14,15,19].

For elastic scattering the relation $\mathbf{k}_f = \mathbf{k}_i + \mathbf{q}$ must be fulfilled, where $\mathbf{q}$ is the scattering vector and $\mathbf{k}_i$ is the wave vector of the incident Bloch wave that scatters into $\mathbf{k}_f$ on the same constant-energy cut. If the initial and final states are non-orthogonal they interfere resulting in a standing wave characterized by a spatial periodicity equal to $2\pi/|\mathbf{q}|$ which is imaged by energy-resolved differential conductivity ($dI/dU$) maps. FT translates the real-space information into reciprocal space, thereby providing direct insight into scattering processes[34].

The typical scattering scenario on TIs, with backscattering forbidden by time-reversal (TR) symmetry, is schematically represented in Fig. 1 a. The situation changes as soon as magnetic adatoms are present on the surface. In this case, electrons can flip their spin during a scattering event, thereby allowing backscattering. Since eigenstates of electrons with opposite spin directions are orthogonal, their initial and final state wave functions cannot interfere (Fig. 1 b). As the concentration of magnetic moments exceeds a certain threshold, they can couple due to the 2D RKKY interaction mediated by the Dirac fermions. Consequently, topological states are expected to acquire a non-vanishing spin component along the magnetization direction, thereby allowing incoming and outgoing electrons to interfere (Fig. 1 c). This corresponds to the moment when TR symmetry breaks, which in FT-QPI is signaled by the appearance of scattering vectors corresponding to backscattering events.

**Transition metal atom adsorption and energy level alignment**

Figure 2 a and b shows two constant-current images taken on the very same sample region before and after Co deposition as spotlighted by arrows pointing at some characteristic defects. The same procedure has been applied for Mn (see Supplementary Figure S1). Figure 2 c, d, and e reports scanning tunneling spectroscopy (STS) data obtained on the pristine, Co-, and Mn-doped $Bi_2Te_3$, respectively. For both Co and Mn the coverage corresponds to approximately 1% of a monolayer (ML). Close inspection of constant-current images reveals two distinct topographical features appearing after the deposition as also observed for other 3d metals[24]. They correspond to single atoms adsorbed in hollow sites of the surface Te layer which are inequivalent with respect to the lattice of the subsurface Bi layer (see Supplementary Figure S2).

The comparison of STS data taken before and after Co or Mn deposition reveals a rigid shift towards negative energies which amounts to approximately 100 meV in either case, indicating a downward band bending that results in an n-doped surface. The position of the valence band maximum and the conduction band minimum was determined following the procedure presented in Ref. 35 (see Supplementary Figure S3). STS spectra obtained by positioning the tip on top of the adatoms can be found in Supplementary Figure S4. Note that for both adatom species, Co and Mn, the Fermi level is energetically located within the bulk gap at this adatom concentration. If there was a gap opening in the topological states, it would be invisible to STS being the Dirac point lower than the valence band maximum, as reported for ferromagnetic bulk doped samples[17,19].

**Spectroscopic mapping of scattering channels−low coverages**

Figure 3 a, b, and c reports energy-resolved QPI maps with the respective Fourier transform (FT) for pristine, Co-, and Mn-doped $Bi_2Te_3$ obtained over the respective regions displayed in Fig. 2. QPI maps obtained on several pristine surfaces always showed the same behavior, indicating the high homogeneity of our crystals. The pristine $Bi_2Te_3$ displayed in Fig. 3 refers to the very same sample region before Co evaporation. Data obtained before Mn deposition can be found in Supplementary Figure S5.

The FT-QPI maps exhibit six-fold symmetric peaks oriented along the Γ−M direction. They correspond to scattering vector $q_2$ (see Fig. 3 d) which connects parallel segments of next-nearest neighbor valleys of the warped equipotential surface and is routinely found on pristine[15] and bulk doped TIs[19,36]. Note, that none of the Fourier-transformed QPI maps in Fig. 3 shows an intensity maximum in the Γ−K direction, indicating that backscattering is absent within the energy range detected here, i.e. far above the Fermi level. Quantitative analysis of these data allows obtaining the energy dispersion relation of the Dirac cone along the Γ−K direction. Results are summarized in Fig. 3 e. A linear fit provides the following values for the Dirac point: $E_D$ = -155 meV, $E_D$ = -245 meV, and $E_D$= -258 meV for pristine, Co-, and Mn-doped $Bi_2Te_3$, respectively.

This scattering scenario profoundly changes for states close to the Fermi level when Mn is present on the surface as reported in Fig. 4. Note that because of the negative energy shift introduced by the surface dopants (see above), comparison of these FT-QPI maps with those obtained on the pristine surface should be done with respect to the panel corresponding to an energy of +100 meV in Fig 2 a. Any change in the scattering scenario and the consequent appearance of new scattering vectors can be thus unambiguously attributed to the effect of the magnetic impurities introduced onto the surface. While no clear scattering channels can be identified once Co adatoms are deposited onto the surface (see Fig. 4 a), a new scattering vector pointing along Γ−K direction appears for Mn (see Fig. 4 b), with an intensity that decreases by moving towards higher energies. Its origin can be understood by direct comparison with the ARPES constant-energy cut reported in Fig. 4 c independently obtained on a $Bi_2Te_3$ sample with a Dirac point located at approximately $E_D$= -250 meV, as in the present case[3]. Its inspection shows that the direction and length of the new feature in our FT-QPI maps correspond to backscattering events $q_1$ [16,19], forbidden by TR symmetry on pristine TIs.

Note that, although in QPI experiments the presence of anisotropic scattering centers may results in additional features appearing in the FT, we can exclude that the additional scattering vector is caused by the shape of the dopants. Indeed, the Mn atoms slightly triangular appearance would result in intensity along Γ−M direction [see supplementary material in Ref. 36], further suppressing the signal along Γ−K. Furthermore, the triangular shape of Co is more pronounced (see Supplementary Figure S2), but here the spots indicative for TR symmetry breaking are absent. Instead, our results indicate that TR symmetry has been broken and that magnetic order is present on the surface even for a Mn concentration as low as 1% ML.

The dispersion of $q_1$ is displayed in Fig. 4 d. A linear fitting of the data results in a Dirac point at $E_D$ = -(263 ± 13) meV, in good agreement with the data of Fig. 3 e. The origin of the magnetic order observed for Mn is attributed to the Dirac fermions which mediate a 2D RKKY interaction among magnetic surface impurities. The oscillation period of the RKKY interaction, which marks the transition from ferromagnetic to antiferromagnetic coupling, is determined by the Fermi wavelength $\lambda_F$. In our experiments $\lambda_F$ = 7 nm, while the average distance between nearest-neighbor adatoms amounts to 3 nm. Therefore, although not homogeneously distributed, the magnetic moments are always ferromagnetically coupled.

**On the importance of magnetic anisotropy**

A detailed theoretical modeling of our experimental findings goes beyond the scope of the present work since it requires a detailed knowledge of the exchange interaction and the constant energy contours. Indeed, it has been recently shown that the effective exchange interaction among single magnetic impurities is reduced by increasing the number of electrons per atom[37]. Furthermore we would like to point out that the evolution from a circle to a convex hexagon may strongly affect the behavior of topological state in the presence of magnetic impurities since it can strongly enhance the strength of the RKKY interaction along some direction as described in Ref. 38. Nevertheless, a simple picture based on an extended Fu model[39,40] which includes the magnetization through a term $H = J \mathbf{S} \cdot \mathbf{\sigma}$, where $J$ is the exchange constant that couples the electron spin $\mathbf{\sigma}$ to the magnetic moment $\mathbf{S}$ introduced by the magnetic adatoms (see Supplementary Note 1) can effectively explain our finding. In particular, they may be a direct consequence of the different Co and Mn single atoms anisotropy axis, which XMCD measurements have shown pointing in-plane and out-of plane, respectively[25,41].

If the magnetization points along the out-of-plane direction, a gap opens at the Dirac point (see dashed line in Fig. 4 e and Supplementary Figure S6). This is not the case for an in-plane magnetization. Figure 4 f and g reports the spin textures at different energies along the magnetization direction for in-plane and out-of-plane directions, respectively (spin texture along the magnetization direction for all directions, i.e. *x*, *y* and *z*, can be found in Supplementary Figure S7). No change in the spin texture is visible for an in-plane magnetization. If magnetic moments are pointing out-of-plane, on the other hand, a spin component along the magnetization direction is introduced while preserving the in-plane spin-momentum locking. This perpendicular spin polarization is particularly strong for energies close to the Dirac point and heals when moving towards higher energies, in agreement with our experimental findings. This trend can be explained by the competition between the ferromagnetic order, which wants to keep all spins aligned along the same direction, and the out-of-plane spin texture imposed by the warping term[40], which forces adjacent valleys to have opposite out-of-plane spin directions.

**Spectroscopic mapping of scattering channels−high coverages**

Interestingly, by continuously increasing the magnetic adatoms concentration, backscattering disappears when the conduction band minimum is moved below the Fermi level as a consequence of the above mentioned adatoms induced band banding. In the present case, this happens at a concentration of approximately 2% ML, at which the conventional scattering scenario on TIs is recovered as shown in Fig. 5. Two different mechanisms can explain the observed behavior. The first is that the presence of bulk states at the Fermi level marks a transition from 2D to a 3D RKKY interaction, for which the coupling strength between magnetic moments decays faster [$R^{-3}$ with respect to the $R^{-2}$ found in 2D systems, being *R* the distance between magnetic moments]. The second is the larger negative shift, which increases the strength of the warping term for topological states located at the Fermi level, imposing its typical texture over that required by ferromagnetic coupling.

This mechanism may also explain why previous reports investigating the interaction of surface magnetic dopants with topological states did not detect any signature of surface magnetic order. Indeed, those experiments have been performed on n-doped samples showing a conduction band lying well below the Fermi level, thus creating a complicating interplay between 2D and 3D states[21-23].

Overall, our findings shed light on the nanoscale interaction of magnetic moments with topological states. The observations reported here show that surface-confined Dirac fermions can drive the coupling between magnetic moments at extremely dilute concentrations establishing local magnetic order. Our results pave the way to explore atomic scale magnetism on TI surfaces and enable, through atomic manipulation or lithographic techniques, a true nanoscale engineering of topological states with arbitrary spin texture where new interesting effects may be observed.

## Methods

### Crystal growth

The Bi$_2$Te$_3$ single-crystals have been grown by a modified Bridgman technique. Bi and Te of 99.999% purity were used for synthesis of polycrystalline Bi$_2$Te$_3$[42]. The starting composition was chosen as 60 mol-% of tellurium that produced the crystals with p-type conductivity and carrier concentration of $10^{18}$ cm$^{-3}$. Suspended components were sealed off in an ampoule, which was pumped down to a residual pressure of $10^{-4}$ Torr. The synthesis furnace was heated at the rate of 50 K per hour up to 870 K, and after 10 hours of melt homogenization, it was switched off. Growth of single crystals was performed in carbon-coated ampoules with a 4 cm tip for geometrical selection of spontaneous microcrystals. The inner diameter of the ampoules was 14 mm, while that of the tip was 4 mm. Crystal growth procedure was carried out with a 10 K cm$^{-1}$ temperature gradient at the front of crystallization. After pulling out the ampoule at the rate of 5 mm per day, the furnace was switched off. After growth, crystals have been cut in sizes suitable for STM experiments (2x2x0.1 mm$^3$).

### STM measurements

The experiments have been performed in an ultra-high vacuum system equipped with a cryogenic STM. After insertion into the system, Bi$_2$Te$_3$ samples have been cleaved at room temperature at a base pressure of $3 \cdot 10^{-11}$ mbar and immediately inserted into the STM operated at $T = 4.8$ K. All measurements have been performed using electrochemically etched tungsten tips. Spectroscopic data have been obtained using the lock-in technique and a bias voltage modulation in between 1 and 10 meV (rms) at a frequency of 793 Hz, with the amplitude progressively increasing with the scanning bias. d$I$/d$U$ maps have been acquired simultaneously to topographic images in constant current mode. Before Fourier transformation, the average value has been subtracted to each d$I$/d$U$ map. To increase the signal to noise ratio, FT d$I$/d$U$ maps have been 3-fold symmetrized according to the rotational symmetry of the underlying surface. Co and Mn (Alfa Aesar) were deposited using an e-beam evaporator with the sample kept at $T = 5.3$ K at an evaporation rate of 1% ML/min. Co and Mn purity was verified by looking at the Kondo state at the Ag(111) surface [43] and at the antiferromagnetic ground state of Mn islands on W(110) [44], respectively.

**Figures**

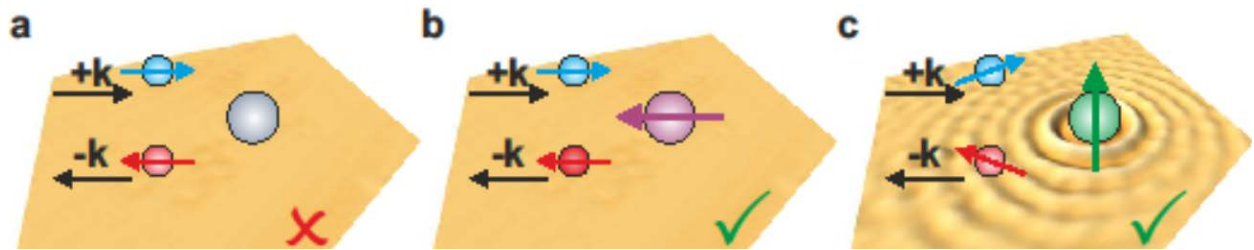

**Figure 1 | Backscattering processes on TIs.** While backscattering events are forbidden by TR symmetry on a pristine TI surface **a**, the presence of magnetic impurities enables spin-flip processes which lead to backscattering **b**. Since, however, orthogonal spin states cannot interfere, backscattering events contribute to the QPI signal only when TR symmetry is broken and magnetic moments become coupled through indirect exchange mediated by Dirac fermions which acquire a non vanishing spin component along the magnetization direction **c**. Black arrows refer to the propagation directions, while colored arrows indicate the spin directions.

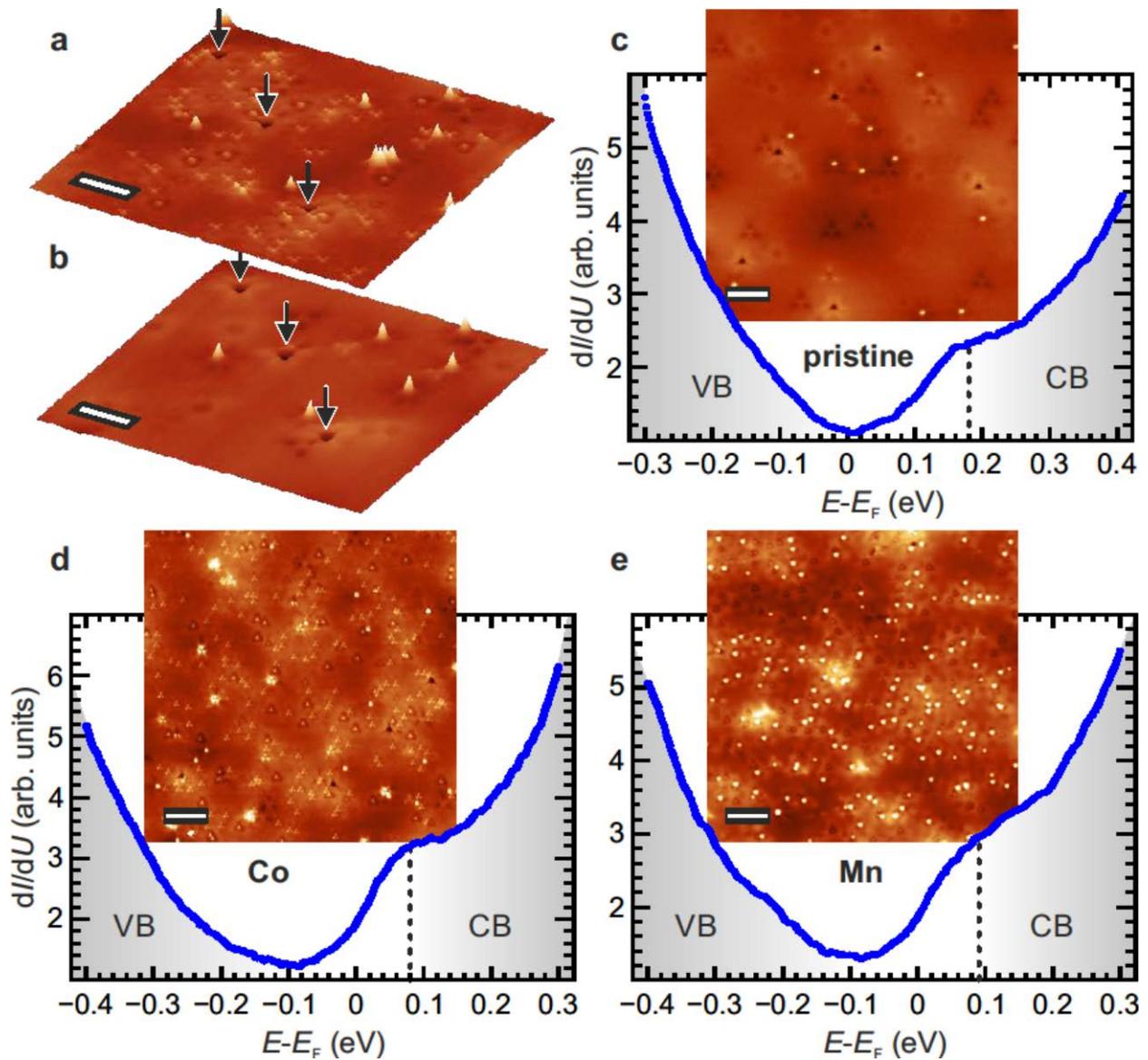

**Figure 2 | Pristine and magnetic surface doped $Bi_2Te_3$.** Constant current images taken on the very same sample region before **a** and after **b** Co atoms deposition. Arrows point at surface defects used as markers. White bars correspond to 5 nm. **c**, **d**, and **e** STS curves taken on pristine, Co, and Mn doped surfaces, respectively. Each inset reports the corresponding constant-current images over which STS spectra have been acquired at off-dopant locations. White bars correspond to 10 nm. Scanning parameters: $I$ = 50 pA, $U$ = -0.3 V.

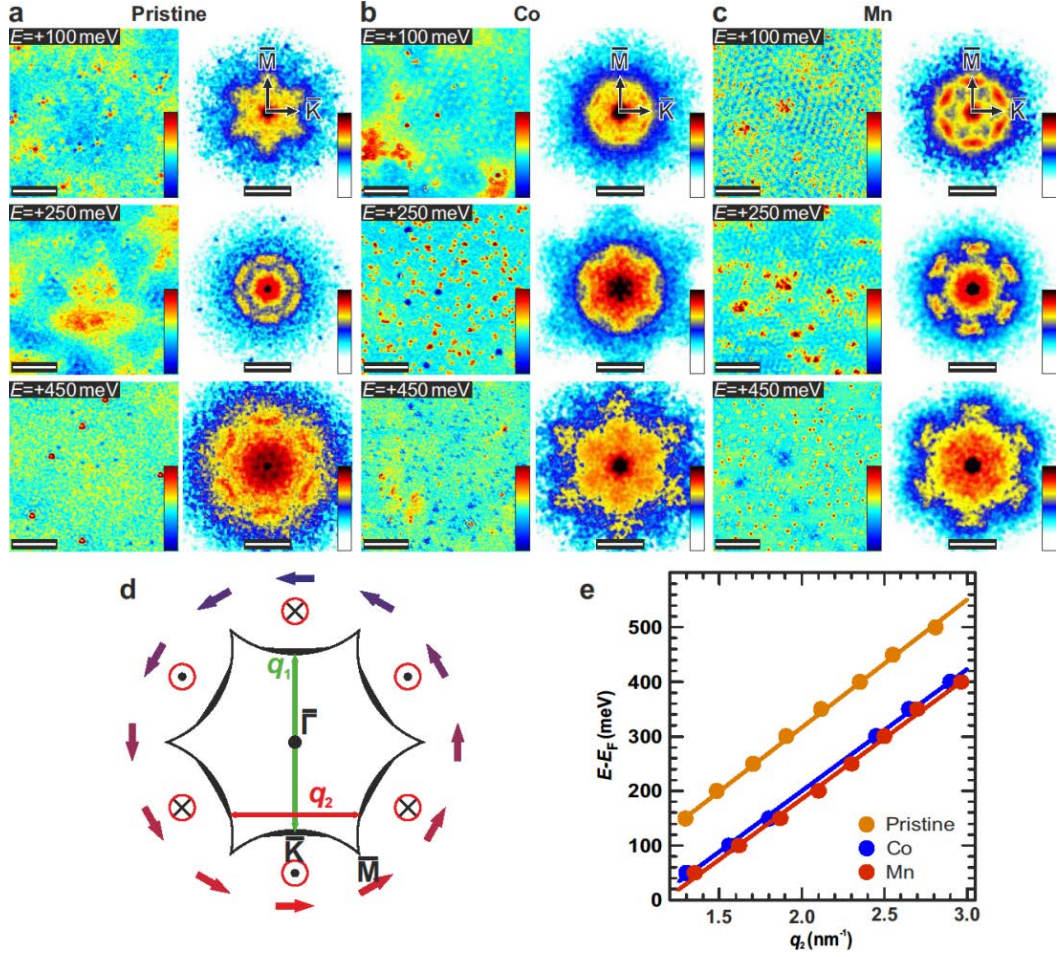

**Figure 3 | Spectroscopic mapping of scattering channels.** d$I$/d$U$ maps with corresponding FT obtained at different energies on the **a** pristine, **b** Co, and **c** Mn surface doped $Bi_2Te_3$. The labelled arrows in the FT indicate the directions of the surface Brillouin zone. The white bars in the d$I$/d$U$ maps correspond to 20 nm, in the FT panels to 2.5 $nm^{-1}$. The intensity of the d$I$/d$U$-signal is represented by the color scale with red (black) and blue (white) referring to high and low signal in d$I$/d$U$ maps (FTs), respectively. **d** Schematic illustration of the scattering scenario on TIs surfaces. The snowflake shape of the constant energy cut is the result of warping effects. Scattering vectors are indicated by $q_1$ and $q_2$. $q_1$ is forbidden in pristine $Bi_2Te_3$ by TR symmetry, while it becomes allowed once magnetic atoms are present on the surface. However, it can be observed by FT-QPI experiments only when the incoming and outcoming electrons have non-orthogonal spin states, i.e. magnetic order is present on the surface (see main text) **e** Energy dispersion of the scattering vector $q_2$, the only one usually observed on TIs surfaces. The plot was obtained by analyzing the FT-QPI pattern at different energies. The straight lines are linear fits to the data.

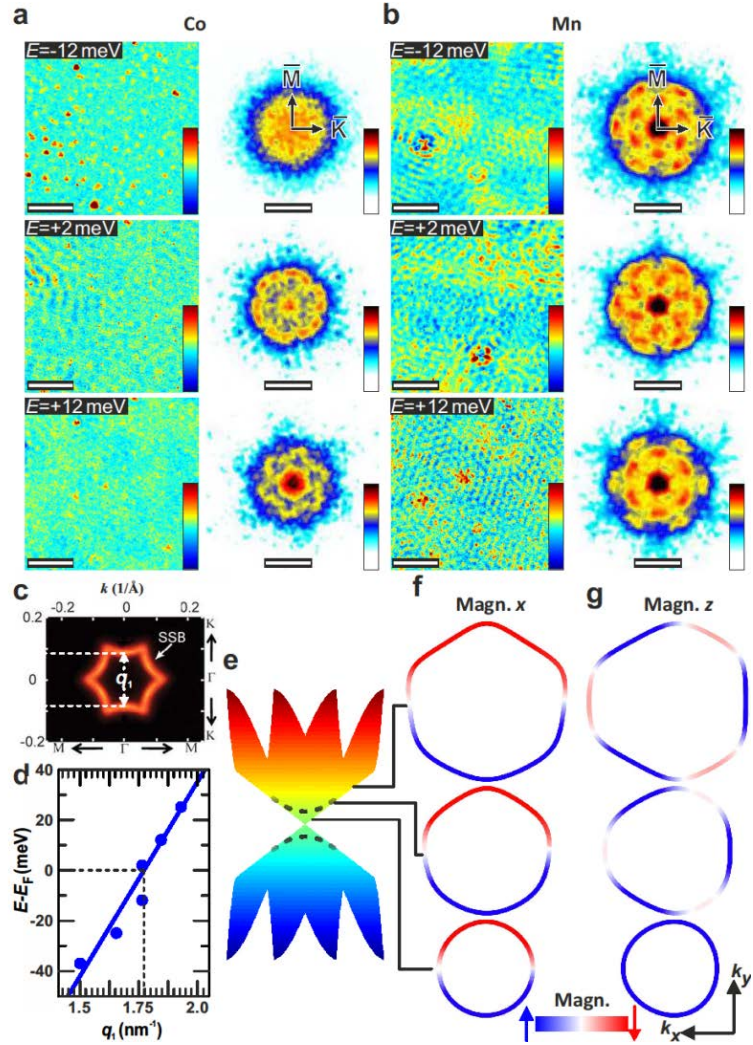

**Figure 4 | Visualizing TR symmetry breaking.** dI/dU maps with corresponding FT obtained at energies close to the Fermi level on **a** Co, and **b** Mn surface doped $Bi_2Te_3$. While no scattering channels are clearly visible for Co, scattering events corresponding to back scattering emerge from the FT QPI patterns in the case of Mn. The white bars in the d*I*/d*U*-maps correspond to 20 nm, in the FT panels to 2 nm$^{-1}$. The intensity of the d*I*/d*U*-signal is represented by the color scale with red (black) and blue (white) referring to high and low signal in d*I*/d*U* maps (FTs), respectively. **c** ARPES constant energy cut obtained at the Fermi level for a $Bi_2Te_3$ sample with a Dirac point located at $E_D = -250$ meV. From [3]. Reprinted with permission from AAAS. **d** Dispersion relation obtained by analyzing the backscattering channels as a function of energy. **e** Band structure and **f**, **g** spin textures along the magnetization direction obtained with an extended Fu model for in-plane and out-of-plane easy axis, respectively. The scale bar refers to the spin polarization along the magnetization direction with blue and red corresponding to opposite spin.

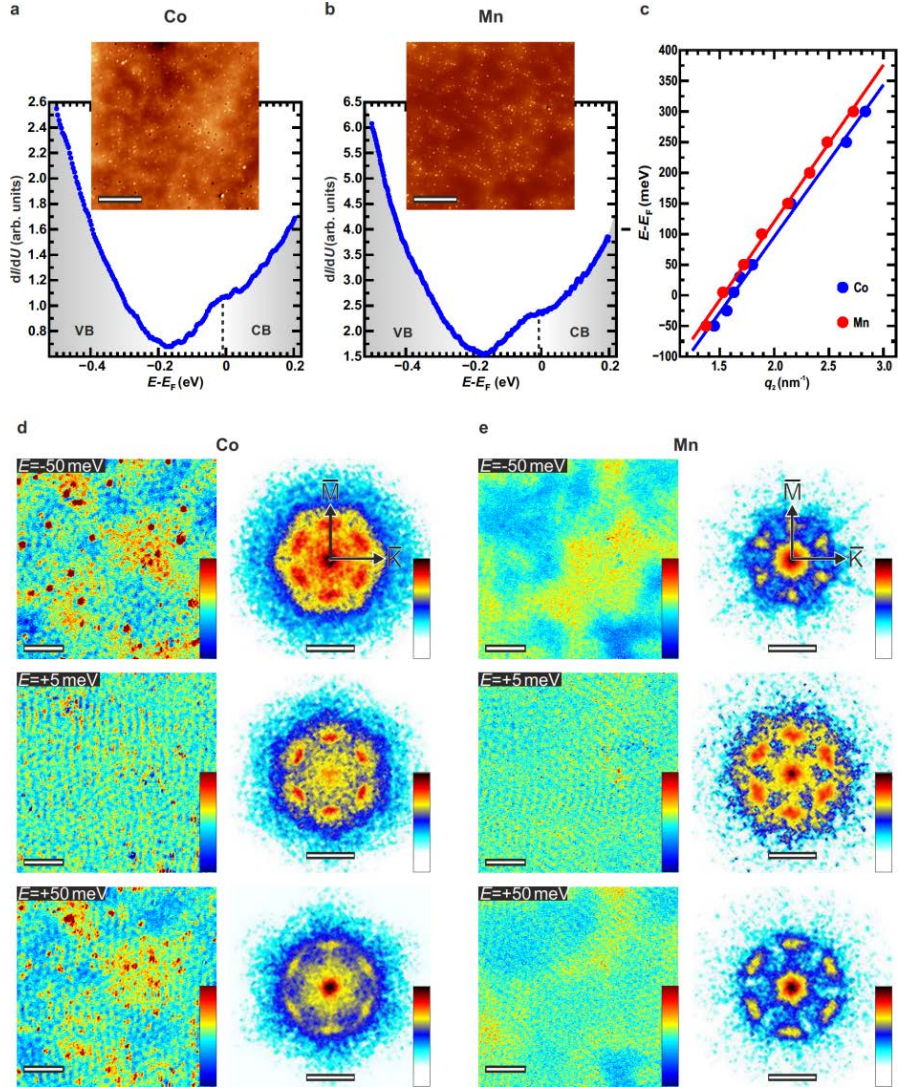

**Figure 5 | Scattering channels at high doping. a**, **b** STS spectra obtained on Co- and Mn- surface doped $Bi_2Te_3$. The adatoms concentration amounts to approximately 2% of a ML. The insets report the corresponding constant-current images where STS have been acquired. The white bars correspond to 20 nm. Note that at this coverage the Fermi levels lies within the bulk conduction band. **c** A linear fit of scattering vectors provides a Dirac point positioned at −393 meV and −385 meV for Co and Mn, respectively. **d**, **e** d$I$/d$U$ maps with their FT for Co and Mn adatoms, respectively. They have been acquired on the same region displayed in the insets. No backscattering vectors appear close to the Fermi level in both cases. The white bars in the d$I$/d$U$-maps correspond to 15 nm, in the FT panels to 2 nm$^{-1}$. The intensity of the d$I$/d$U$-signal is represented by the color scale with red (black) and blue (white) referring to high and low signal in d$I$/d$U$ maps (FTs), respectively.

**Supplementary Figures**

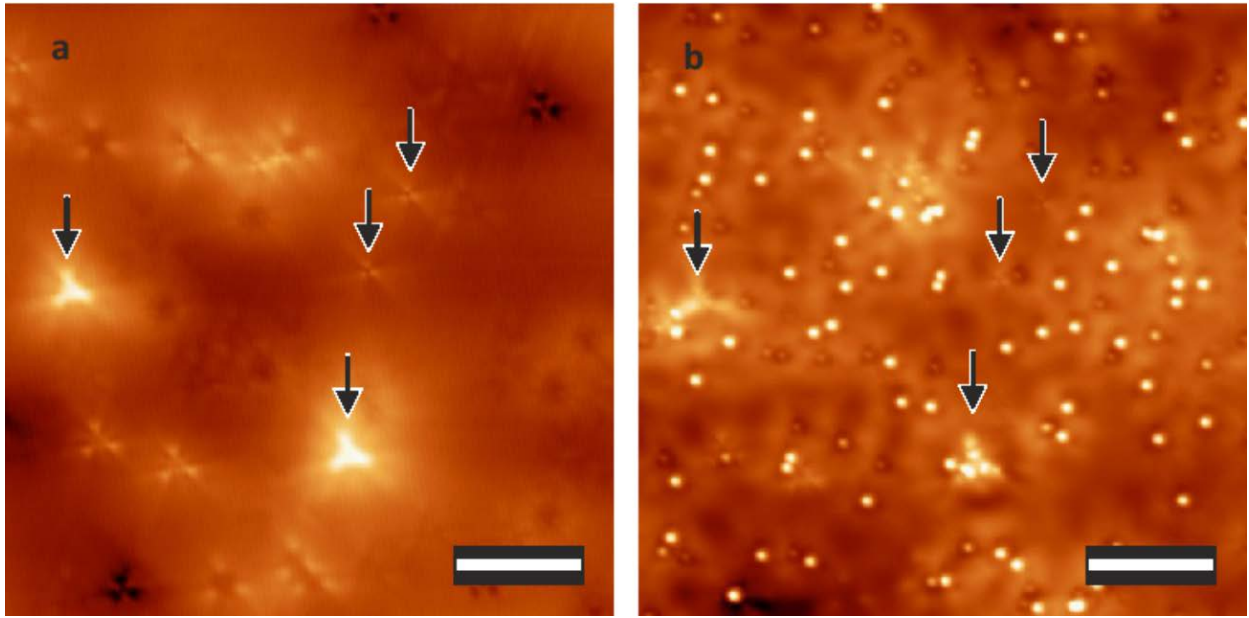

**Supplementary Figure S1: Region mapping**. **a** Pristine and **b** Mn-doped $Bi_2Te_3$. Arrows point at characteristic defects present on the pristine surface which have been used as markers to map the very same sample region before and after deposition. The white bars correspond to 10 nm.

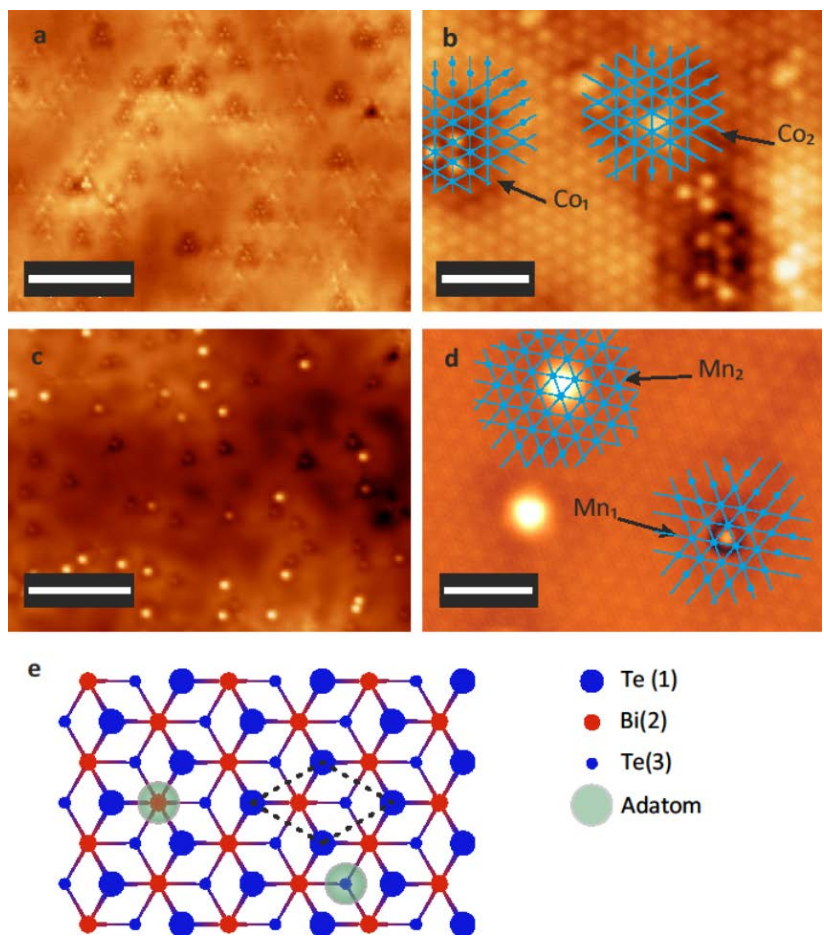

**Supplementary Figure S2: Adsorption sites for Co and Mn adatoms. a**, **b** and **c**, **d** show the adsorption sites for Co and Mn atoms, respectively. For both elements two adsorption sites have been identified. They correspond to hollow sites of the surface Te layer which are inequivalent with respect to the lattice of the subsurface Bi layer as illustrated in **e**. Because of the lack of subsurface resolution, an unambiguous assignment of these two feature to hcp and fcc hollow sites is not possible. In both cases the corrugation of the adatoms is found to be small, spanning a range of a few tens of pm depending on the scanning parameters. This suggests a strong inward relaxation, consistent with the relative small size of 3d elements compared to the substrate inter-lattice spacing. The same has been reported to occur for Fe and Co atoms on $Bi_2Se_3$[1,2]. White bars in a and c correspond to 10 nm, in b and d to 2 nm.

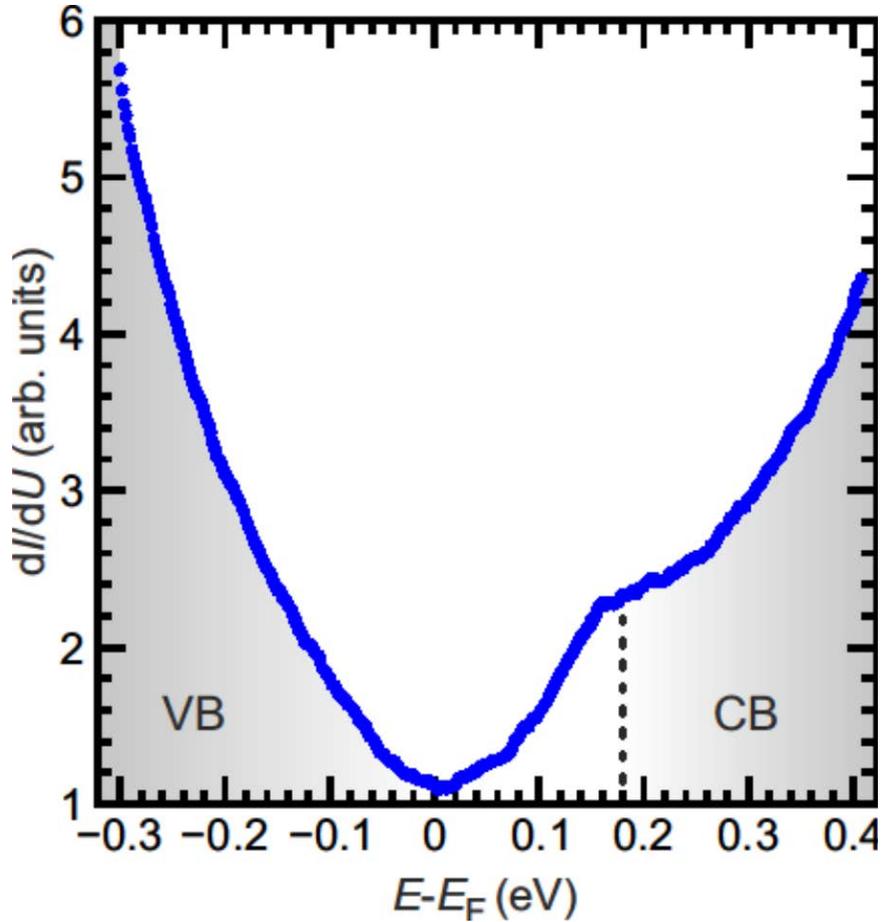

**Supplementary Figure S3: Assignment of bulk bands.** STS spectrum obtained on pristine $Bi_2Te_3$. The minimum close to the Fermi level and the shoulder visible at approximately 180 meV correspond to the valence band (VB) maximum and conduction band (CB) minimum, respectively. Assignment of this features was done according to the procedure described in Ref. 3, 4. Compared to Ref. 4, these features are shifted in energy because of different sample doping. Scanning parameters: $I = 50$ pA, $U = -0.3$ V.

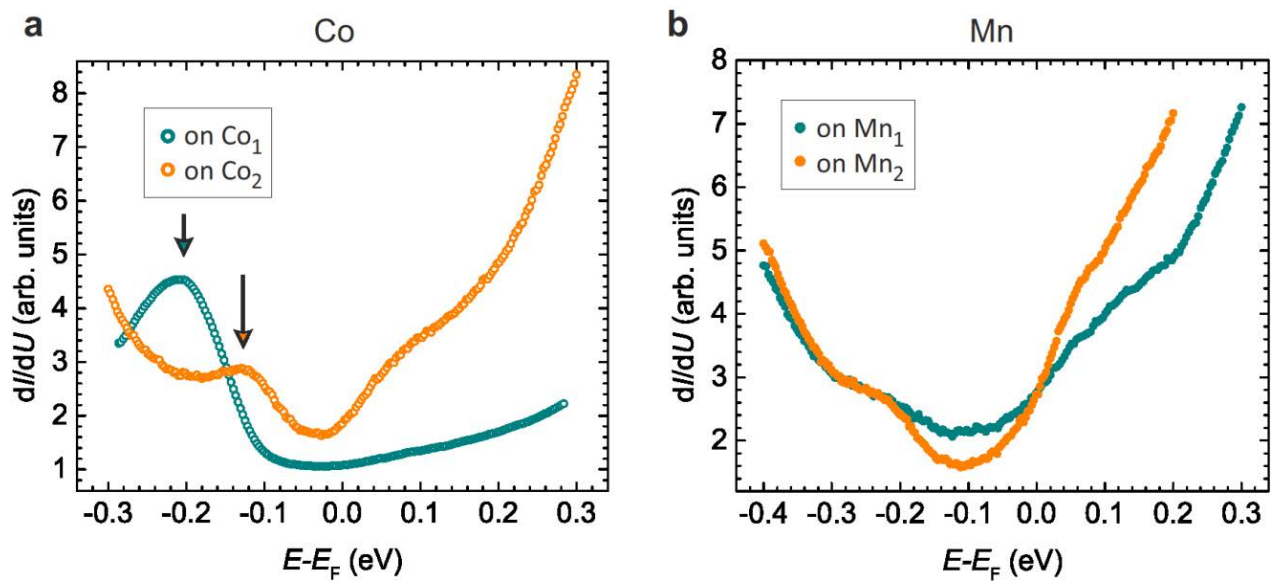

**Supplementary Figure S4: STS on adatoms. a**, **b** show STS spectra obtained by positioning the STM tip on top of the Co and Mn adatoms, respectively. $Co_1$ and $Co_2$ as well as $Mn_1$ and $Mn_2$ refer to the adsorption position. While resonances corresponding to impurity states consistent with theoretical predictions[5] are detected for Co (see black arrows in a), no resonances have been detected for Mn adatoms in energy range of interest.

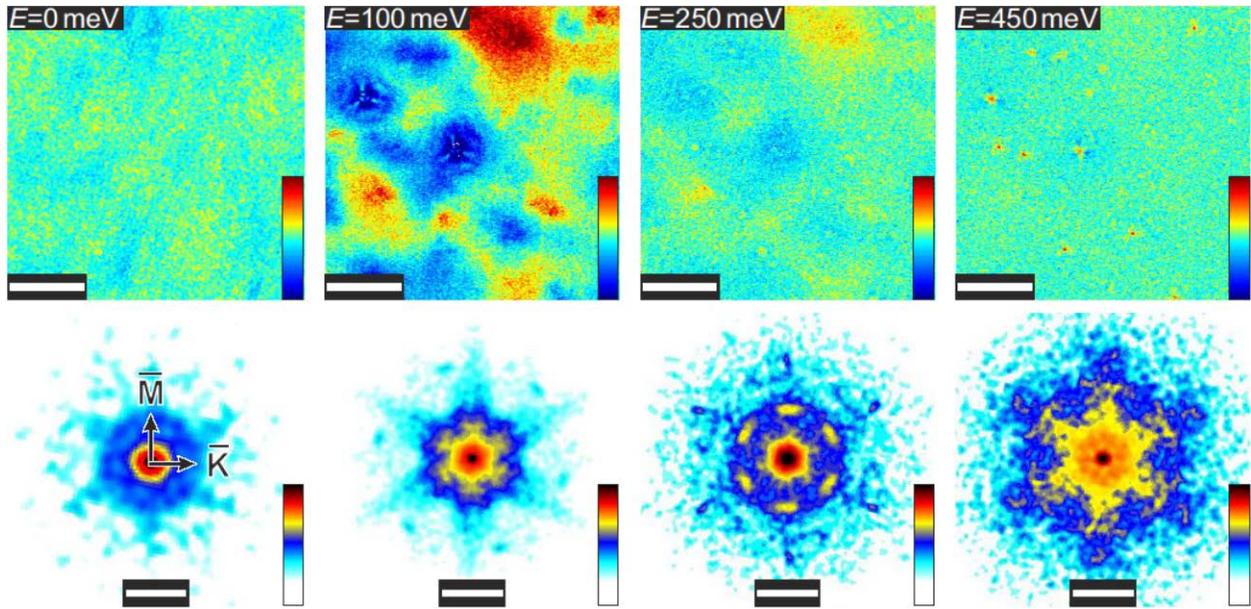

**Supplementary Figure S5**: **FT-QPI on the pristine sample before Mn deposition.** d$I$/d$U$ maps and their relative FT obtained over the pristine Bi$_2$Te$_3$ region diplayed in Supplementary Figure S1 before Mn deposition. Note that there is no indication for TR symmetry breaking (no scattering intensity in Γ−K direction). White bars in d$I$/d$U$-maps correspond to 10 nm, in FT to 2 nm$^{-1}$. The intensity of the d$I$/d$U$-signal is represented by the color scale with red (black) and blue (white) referring to high and low signal in d$I$/d$U$ maps (FTs), respectively.

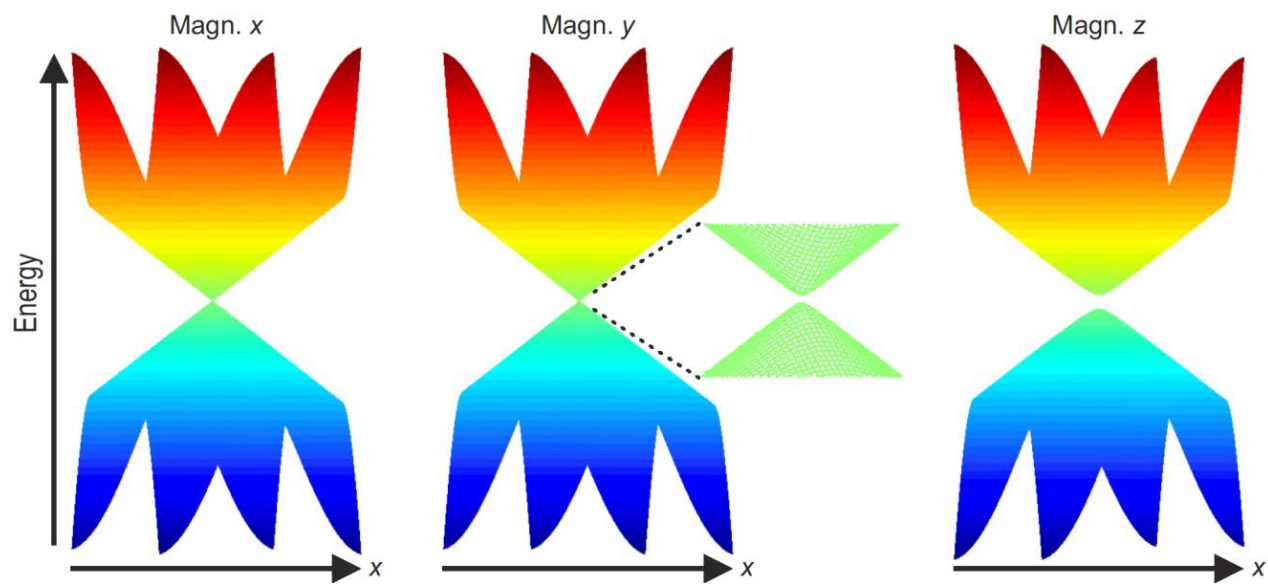

**Supplementary Figure S6: Band structure obtained with an extended Fu model.** Effect on the band structure of $Bi_2Te_3$ for magnetization pointing along *x*, *y* and *z*.

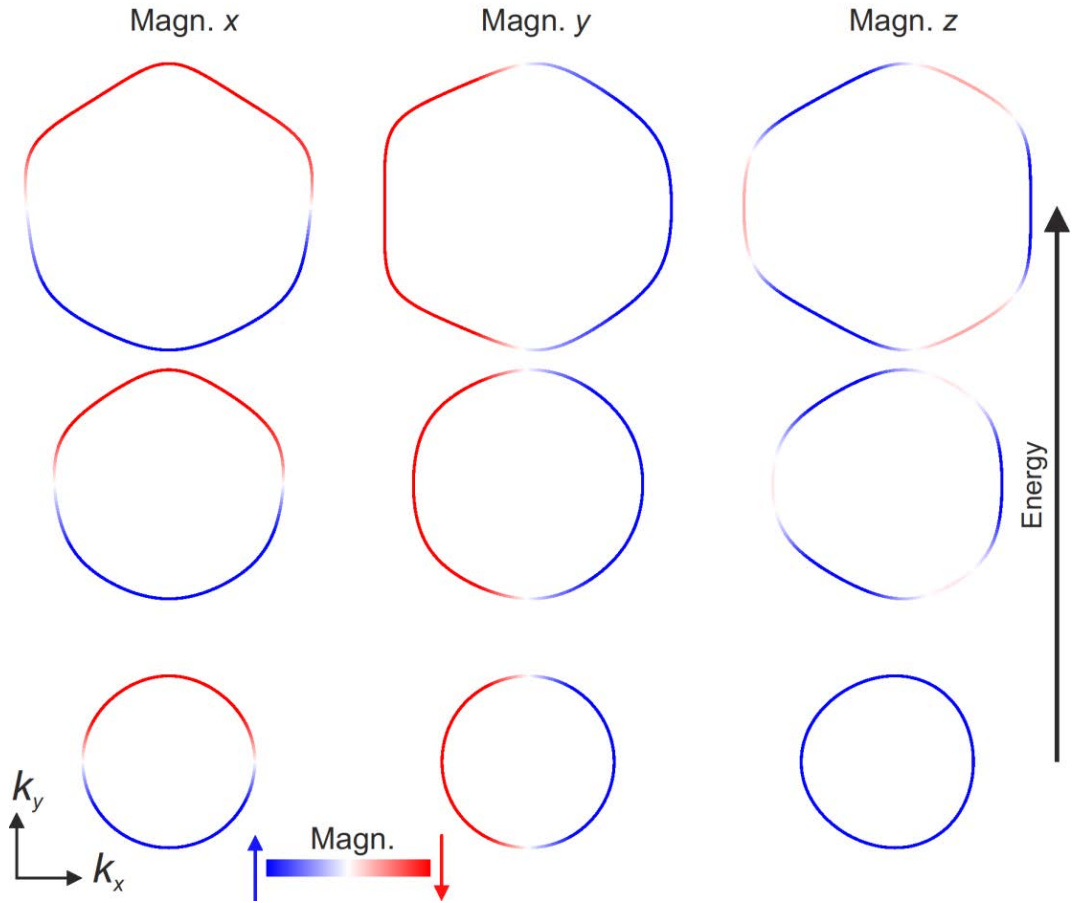

**Supplementary Figure S7: Spin texture along different magnetization directions.** A magnetization of the topological states takes place only for an out-of-plane direction. By moving at higher energies the magnetization progressively disappears as the warping term increases its strength. The scale bar refers to the spin polarization along the magnetization direction with blue and red corresponding to opposite spin.

**Supplementary Note 1**

**Modeling magnetic impurities**

The influence of the magnetic moments present on the surface of $Bi_2Te_3$ has been described by an extended Fu model[6]:

$$H = v_k \cdot ( k_x \sigma_y - k_y \sigma_x ) + \lambda ( k_+^3 + k_-^3 ) \sigma_z - J\, \mathbf{S}\cdot\boldsymbol{\sigma},$$

where $v_k$ is the velocity, $k_\pm = (k_x \pm i k_y)$ and $J\,\mathbf{S}\cdot\boldsymbol{\sigma}$ models the interaction of the electron spin $\boldsymbol{\sigma}$ with the moment $\mathbf{S}$ introduced by the magnetic atoms. As shown in Supplementary Ref. 7, such a model can reproduce the results obtained by *ab-initio* calculations. Its simplicity helps to understand the properties of topological states interacting with magnetic moments. Supplementary Figure S6 shows the effect of a magnetic term pointing along *x*, *y*, and *z* directions. *x* and *y* are two inequivalent in-plane directions which result from the C3v symmetry of the crystal, which consists of a threefold symmetry along the direction *z* perpendicular to the surface and a reflection at the *yz* plane for which the following relation holds: *x* → -*x* with *x* pointing along the Γ-K direction.

A gap opening at the Dirac point is clearly visible only when the magnetization is pointing out-of-plane, but not for in-plane magnetizations. Note that the *x* and *y* directions are not equivalent. While the Dirac point stays intact for a magnetization pointing along *x*, a tiny gap (approximately 80 times smaller than that observed along *z*) opens along *y* as highlight in the inset. This is the result of the above mentioned symmetry operations: a magnetic term pointing along *x* preserves mirror symmetry on the *yz* plane while one pointing along *y* does not.

The resulting spin texture of the topological state is illustrated in Supplementary Figure S7. An in-plane magnetization leaves the spin texture essentially identical to the pristine case. On the other hand, for a magnetization pointing along *z*, the spin texture is substantially altered and a magnetization of the topological state appears, which progressively vanishes by moving towards higher energies, where the strength of the warping term increases.